\documentclass[showpacs,reprint,preprintnumbers,amsmath,amssymb,superscriptaddress,longbibliography,nofootinbib]{revtex4-2}

\usepackage[utf8]{inputenc}
\usepackage{amsmath}
\usepackage{amssymb} 
\usepackage{enumerate}
\usepackage{multirow}
\usepackage{braket}
\usepackage[toc,page]{appendix}
\usepackage{url}
\usepackage{graphicx}
\usepackage[pdftex,bookmarks,linktocpage,pdfpagelabels,plainpages=false,hyperfigures,linkcolor=blue,citecolor=blue]{hyperref} 
\hypersetup{colorlinks=true}
\usepackage{adjustbox}
\usepackage{mathrsfs,amssymb}  
\usepackage{cancel}
\usepackage[normalem]{ulem}
\usepackage{cleveref}
\usepackage{tikz}
\usetikzlibrary{decorations.markings}
\usetikzlibrary{positioning}
\usetikzlibrary{shapes}
\usetikzlibrary{calc}

\newcommand{\beq}{\begin{equation}}
\newcommand{\eeq}{\end{equation}}

\newcommand{\bea}{\begin{eqnarray}}
\newcommand{\eea}{\end{eqnarray}}

\newcommand{\GeV}{{\rm GeV}}

\begin{document}

\title{Quantum Metric Learning for New Physics Searches at the LHC } 
\author{A.\,Hammad}
\email{hamed@post.kek.jp}
\affiliation{Institute of Convergence Fundamental Studies, Seoultech, Seoul 01811, Korea}
\affiliation{Theory Center, IPNS, KEK, 1-1 Oho, Tsukuba, Ibaraki 305-0801, Japan}
\author{Kyoungchul Kong}
\email{kckong@ku.edu}
\affiliation{Department of Physics and Astronomy, University of Kansas, Lawrence, KS 66045, USA}
\author{Myeonghun Park}
\email{parc.seoultech@seoultech.ac.kr}
\affiliation{Institute of Convergence Fundamental Studies, Seoultech, Seoul 01811, Korea}
\affiliation{School of Natural Sciences, Seoultech, Seoul 01811, Korea}

\author{Soyoung Shim}
\email{soyoung@kaist.ac.kr}
\affiliation{Institute of Convergence Fundamental Studies, Seoultech, Seoul 01811, Korea}
\affiliation{School of Electrical Engineering, Korea Advanced Institute of Science and Technology (KAIST), 291 Daehak-ro, Yuseong-gu, Daejeon 34141, Republic of Korea}
\date{November 29, 2023}

\begin{abstract}
In the NISQ (Noisy intermediate-scale quantum) area, Quantum computers can be utilized for deep learning by treating variational quantum circuits as neural network models.
 This can be achieved by first encoding the input data onto quantum computers using nonparametric unitary gates. 
 An alternative approach is to train the data encoding to map input data from different classes to separated locations in the Hilbert space.
The separation is achieved with  metric loss functions, hence the naming ``Quantum Metric Learning".  With  the limited number of qubits in the  NISQ area, this approach works naturally as a hybrid classical-quantum computation enabling embedding of high-dimensional feature data into a small number of qubits. Here, we consider an example of the global QCD color structure of hard b-jets emerging from color singlet scalar decays to optimize the signal to background discrimination with a hybrid classical-quantum metric learning. Due to the sparsity of data, self-supervised methods with data augmentation have been utilized so far. Compared to the this classical self-supervised approach, our hybrid method shows the better classification performance without data augmentations. We emphasize that performance enhancements independent of data augmentation techniques are devoid of the artificial risks introduced by data augmentation.
 \end{abstract}

\maketitle
\section{\bf Introduction}  
Quantum machine learning (QML) has recently found applications in High Energy Physics (HEP) for various tasks \cite{Blance:2020nhl,Terashi:2020wfi,Chen:2020zkj,Wu:2020cye,Gianelle:2022unu,Belis:2021zqi}. The strength of QML lies in the capability of quantum computers to coherently store and process extensive data within the tensor product Hilbert space. Quantum computations have demonstrated a significant speedup in HEP analysis, particularly with small training data sets \cite{Guan:2020bdl,Alvi:2022fkk}. One specific implementation is the Variational Quantum Classifier (VQC), a type of quantum neural network suitable for both supervised and unsupervised learning. This is achieved by constructing a quantum circuit parameterized by a set of tunable parameters. Training a VQC closely resembles classical machine learning: after encoding the data onto a quantum computer, the VQC is optimized to minimize a loss function by adjusting the free parameters.

A critical aspect of VQC design is the data encoding strategy, considered as a quantum kernel mapping of classical data onto the Hilbert space spanned by the qubits \cite{havlivcek2019supervised,schuld2021effect}. 
Traditional data encodings, such as amplitude and angle encoding, linearly map the input data to the registered qubits.
However, for non-linearly separable data, both kernels prove inefficient and limit the expressive power of the VQC. An alternative encoding method involves using a non-linear kernel mapping, achieved by repeating the data encoding after each layer of the VQC \cite{perez2020data}. In this manner, it is possible to train the data encoding in such a way that the distance between the encoded data of different classes is maximized, hence the term ``Quantum metric learning." Indeed, quantum metric learning can be viewed as the quantum counterpart to classical contrastive learning models \cite{chen2020simple,grill2020bootstrap,caron2020unsupervised}.    

Although the classical contrastive learning models can efficiently deal with unsupervised classification tasks, they suffer from two main problems, the collapse of the projection heads \cite{jing2021understanding} and the impact of data augmentation on the classification performance \cite{zhang2022rethinking}.  Data augmentation and global weights sharing are the core points in classical contrastive learning to learn the similarity or dissimilarity between inputs from different classes. The impact of augmentations on the input data  enhances the classification performance by generalizing the boundaries of the mapped data  from different classes into the latent space of the model.  In some HEP analysis, treating each type of augmentation equally during training makes the model learn non-optimal features  and limits the flexibility to choose augmentation types with optimal  features beforehand. Moreover, 
excessive invariance induced by overly strong data augmentations in traditional contrastive learning methods can lead to the loss of fine-grained information essential for certain downstream tasks.


Quantum metric learning successfully addresses these issues by preventing the collapse of projection heads, achieved through the quantum kernel mapping of data onto the high dimensional Hilbert space of the entangled qubits. Notably, the mapping of classical data onto a high dimensional Hilbert space is a characteristic feature of any  (VQC) with a larger number of entangled qubits, independent of the data encoding type. To enable a VQC to learn the similarity of input data, a distance-based loss function is required. The simplest approach involves minimizing the overlap between encoded data in a supervised manner, as demonstrated in \cite{perez2020data}. In this scenario, to minimize the loss function, the model enhances the purity of the measured states  by mapping the encoded data  onto different bases of the readout qubit based on the truth labels.
Another method is to maximize the trace distance $(l_1)$ between encoded data from different classes, as in Helstrøm minimum error measurement \cite{helstrom1976quantum}. Alternatively, maximizing the Hilbert-Schmidt (HS) distance $(l_2)$ involves measuring the overlaps between encoded data, such as through a simple swap test, as explored in \cite{lloyd2020quantum}. In this case, the VQC maximizes the purity of the measured state, minimizing the overlap between them without labels. In fact, to maximize the HS distance the VQC pushes the embedded data from different classes into different orthogonal bases of the readout qubit. Consequently, this setup eliminates the necessity for data augmentations to learn the similarity or dissimilarity between input data.

In this work, we take a concrete example of probing the signatures of a CP-odd (A) scalar at the LHC in  the Two Higgs Doublet Model. We consider the process of $A$ decays to $Z$ boson and SM-like Higgs $h$ followed by  $Z \to \ell \bar \ell, h\to b \bar b$. The presence of  two resolved b-jets in the final state offers a peculiar QCD color flow structure with respect to the background. This structure can be seen at the LHC as a  string from the soft hadrons that stretches between two color-connected b-jets \cite{Maltoni:2002mq,Hagiwara:2010vk,Kilian:2012pz}. Specific to our analysis, the two b-quarks from the signal process, which emerged from the Higgs boson decays, form a  color dipole with a tendency for more soft radiation to occur in the region between them. On the other hand, the b-quarks from the background process, which emerged from a QCD colored mediator, form isolated poles that are connected to the parent particle. In fact, this  dynamics can be used as a signature to aid the search for new physics.  
Here, we use the QCD color structure to distinguish the signal events from the background using classical and quantum metric learning models. For this purpose, we first adjust a classical  model in \cite{Dillon:2021gag} which is based on a simple contrastive learning method introduced in \cite{khosla2020supervised}. Moreover, we investigate the effect of different data augmentation, e.g. rotation, translation, smearing, etc, on its classification performance. Also, we adopt a quantum metric learning model with a single qubit that is introduced in \cite{lloyd2020quantum}. To exploit the quantum interference of the entangled qubits we enlarge the depth of the quantum circuit to four registered qubits. 

The structure of the paper is organized as follow,  in the first section we discuss the basics of VQC training followed by  a discussions about the variational data encoding. In the second section  the physics analysis using quantum and classical contrastive learning models . Finally, we present our results and draw our conclusion.
\section{\bf Quantum Machine Learning}

\subsection{\bf Basics of Training a Variational Quantum Circuit}  
As mentioned previously, a VQC can identify features of the classical inputs by  encoding the  input data onto a quantum computer and  apply  unitary transformations on them, while the predicted probability is obtained  by projecting the quantum state on one of the computational bases of the qubit.
Training a VQC 
consists of
the following steps:
\begin{itemize}
\item Prepare quantum states by encoding classical data onto a quantum computer.
\item Apply unitary transformations to the encoded data using quantum variational gates.
\item Measure the expectation value from the readout qubit.
\item Utilize a classical optimizer to derive new values for the quantum variational gates.
\item Iterate through the above steps until the loss function reaches a minimum.
\end{itemize}
Data encoding can be achieved by using a  unitary transformation as 
\begin{equation}
    x \mapsto U(x)|0\rangle^{\otimes n} = |x\rangle \,,
\end{equation}
where $n$ is the number of the used qubits and $U$ is a unitary operator. The parameterization of the encoding, and hence $U(x)$, can affect the decision boundaries of the final predictions and can therefore be chosen in a form that suits the problem at hand \cite{farhi2018classification,Blance:2020nhl}. There are several types of parameterization, e.g. basis encoding, amplitude encoding and angle encoding. These encoding methods map the input data onto the Hilbert space of the qubit based on  fixed rotations of the qubit. A variational encoding is obtained  by training the kernel mapping such that the  distance  between different input data is maximal. Another  encoding is  the  incremental encoding \cite{Periyasamy:2022lzp}, that  maps high-dimensional  data onto VQC with a small number of qubits. After data encoding and  once the quantum states are prepared, the VQC  maps  the prepared states to another states via a set of unitary transformations, $|\psi(x,\theta)\rangle = U(\Vec{\theta})|x\rangle$, with $\Vec{\theta}$ are the tunable parameters to minimize the error between the model predictions and the true values. In general, the unitary transformations of the prepared states can be decomposed of a series of unitary gates as
\begin{equation}
    U(\Vec{\theta}) = U_j(\Vec{\theta}_\sigma) \ldots U_3(\Vec{\theta}_\gamma) U_2(\Vec{\theta}_\beta) U_1(\Vec{\theta}_\alpha) \,,
\end{equation}
where $j$ is the maximum number of unitary gates in the quantum layer. For a VQC with larger number of qubits, the structure of $U$ is decomposed of unitary rotation gates and entangled gates. Common gates that entangle two qubits are the Controlled-Not or Controlled-Z  gates which can be accessed with out tunable parameters. Both gates flip the state of a target qubit based on the value of another control one. Similar to the classical neural network, enlarging the number of the entangled qubits increases the network space that can efficiently express the structure of the data.    After applying the unitary operations, the expectation value of the quantum state can be measured using the Pauli-Z operator as
\begin{equation}
    f_\theta(x) = \langle\psi(x,\theta)|\sigma_Z| \psi(x,\theta)\rangle \,,   
\end{equation}
where $\sigma_Z$ is the Pauli-Z operator.
VQC optimization is done by  minimizing a loss function between the eigenvalues of the measured quantum state and the true labels. Here, as in the classical machine learning methods, minimization of the loss function can be done using a classical optimizer. The back propagation and parameters update is computed using the parameter shift rule \cite{bergholm2018pennylane}
\begin{equation}
    \frac{\partial}{\partial\theta} \mathcal{F} = \frac{\mathcal{F}(\theta+\delta)-\mathcal{F}(\theta-\delta)}{2} \,,
\end{equation}
where $\mathcal{F}$ is the quantum circuit and $\delta= \frac{\pi}{2}$. Once the loss function is minimized we can test the VQC accuracy for new unseen test data similar to the classical neural network.

\begin{figure*}[t!]
    \centering
    \hspace*{-0.4cm}
    \includegraphics[width=1.02\textwidth]{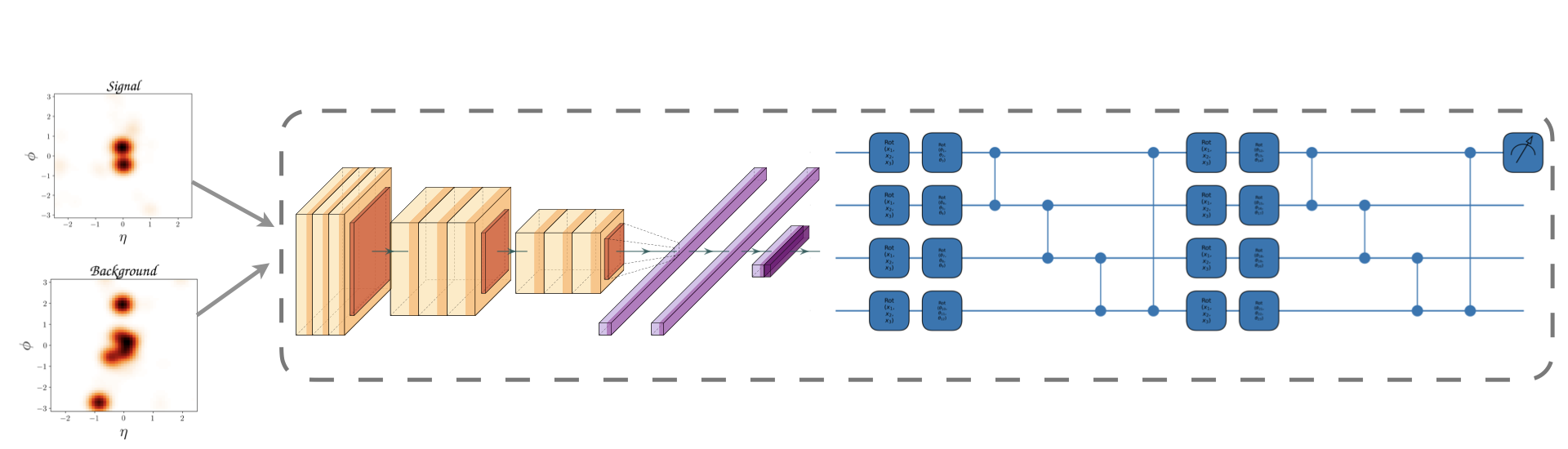}
    \caption{Schematic architecture of the hybrid classical-quantum model.  The dimensions of the input images are reduced via the classical encoder before it passed to the VQC. After measuring the overlap between the quantum states we use a classical optimizer to update the weights of the classical encoder and VQC at each iterations. In our study, we repeat the layers of the VQC 10 times for the case of a single qubit usage and 6 times when we use multi-qubits VQC.  }
    \label{fig:QML}
\end{figure*}
\subsection{\bf Quantum metric learning} The \textit{``traditional"} quantum  learning models, as mentioned above, rely on encoding some classical data $x$ onto quantum state $ |\psi(x)\rangle$, then, a decision boundary is learned by training the variational circuit to adapt the measurement basis.  A classical optimizer can be used to tune the VQC parameters to improve the classification boundary, which is analogous to the classical approach where a decision boundary is learned to separate classes. Instead of focusing on training the VQC to adapt the measurement, quantum metric learning trains the encoding of the data onto the qubit.

Interpreting the data encoding strategy as a feature map reveals that it is more than just a necessary  preparation step in a quantum machine learning algorithm. The reason is that the feature map changes the structure of classical data in a non trivial manner as the encoding gates act as non linear operation on the data. Nonlinear transformation can change the distances between the mapped classical data onto the Hilbert-space spanned by the qubits. In principle, if the mapped data onto the qubit are linearly separable enhances the expressive power of the VQC ending up with an improved classification performance. Traditional training of the  quantum gates acting on quantum states are unitary, and hence linear operations which do not change the distance between the states from different classes. On the other hand, VQC with variational encoding is a novel method to train the mapping of the classical data onto the qubit such that the distance between data from different classes is maximal. 

\subsection{\bf Variational data encoding}
The fact that quantum gates acting on quantum states are unitary does not change the distance between two states. This means that after the data encoding the VQC acts linearly on the embedded  data, while the only source of nonlinearity  comes from the measurement of the quantum state. If the data encoding is not rich enough  the quantum circuit ends up with very limited  expressive power, even if the VQC is arbitrarily deep.
To make the VQC an expressive model, rich embedding and deep quantum circuit are needed. Accordingly, repeating the data encoding layer after each unitary variational layer can achieve the aforementioned tasks. Repeated data encoding and unitary variational layers   can be written as  a sum of a multi-dimensional partial Fourier series \cite{schuld2021effect}
\begin{equation}
    f_\theta(x) =\sum_{\omega=-L}^L c_\omega(\theta)e^{i\omega x} \,,
\end{equation}
where circuit frequency spectrum $\Omega$  ranges between $[-L,L]$ with $L$ is the number of the repeated blocks. The frequency spectrum  of the circuit is   determined by the eigenvalues of the data encoding gates, while its Fourier coefficients depend on the entire circuit. Since quantum models have real valued outputs, the learned function can be expressed as linear combinations of sine and cosine functions of the form $\cos(\omega\ x_i)$ and $\sin(\omega\ x_i)$.  As a matter of fact, the representation of quantum models as Fourier  sums has the property that the frequency spectrum determines the exponential part which  describes the  accessible 
 function by the quantum circuit, while the Fourier coefficients determine how the accessible functions can be combined. These  properties can give insight into the function classes that the quantum model can  learn. In general, one can think of a VQC with repeated encoding as Fourier sum of trigonometric functions with frequencies determined by the data embedding blocks and a combined amplitude determined by the trainable blocks. As far as the circuit ansatz is repeated, the nonlinearity of the learned function is increased, allowing for the high levels of expressivity which is needed to enhance the classification performance.


\subsection{\bf A single qubit classifier} 
A single qubit has the freedom to construct a VQC with repeated data encoding. The structure of the single qubit circuit is define as  
\begin{equation}
    U(\Vec{\theta},\Vec{x}) = \mathcal{L}(N) \ldots \mathcal{L}(2)\mathcal{L}(1) \,,
\end{equation}
where $\vec{\theta} = (\theta_1,\theta_2,\theta_3)$ and $\vec{x} = (x_1,x_2,x_3)$ are the trainable parameters and the input vector to the quantum circuit respectively; while the processing layers are decomposed as $\mathcal{L}(i) = U(\vec{x})U(\vec{\theta}_i)$. An alternative approach is to consider the processing layer as $\mathcal{L}(i) =\Pi^{n}_{i=1} \sigma_x(x_i)\sigma_y(\theta_i)$, with $\sigma_x, \sigma_y$ are Pauli-X and -Y respectively, and $n$ is the dimension of the input data. Although the structure of the processing layer is different, it encodes the input data equally while the only difference is that we can encode higher dimensions data,  $d > 3$, in the latter processing layer. As the VQC with variational encoding, $U(\Vec{\theta},\Vec{x})$, is used to train the encoding such that the distance between input from different classes is maximal, we want to enforce the measured quantum states  to be as near as possible to one particular state on the Bloch sphere by training the model to minimize the overlap between the measured states from different classes. Considering the classification problem with two classes, the computational bases of the readout qubit can be used to cluster data from different class in each basis, $|0\rangle$ or $|1\rangle$. For a  supervised training, the true labels are used to cluster the data from each class on each computational basis by minimizing a fidelity based loss function as \cite{perez2020data}
\begin{equation}
    \mathcal{L}(\vec{\theta},\vec{x}) = \frac{1}{M}\sum^M \left( 1- \left|\langle \psi_c|\hat{\psi}(\vec{\theta},\vec{x})\rangle\right|^2 \right) \,,
    \label{loss:fedal}
\end{equation}
where $M$ is the number of the training data set and $\psi_c$ is the true "pure" states with $c$ indicates either $|0\rangle$ or $|1\rangle$;  $\hat{\psi}(\vec{\theta},\vec{x})$ is the measured state.  The quantity, $\langle \psi_c|\hat{\psi}(\vec{\theta},\vec{x})\rangle $,  defines the fidelity between the measured and the pure state which measures the overlap between them.  In fact, the loss function can be considered as  a metric function as it maximizes the purity of the measured states of each class which in turns  maximizes the  distance between them, i.e. the purity  of the measured states is maximal  by pushing them into two orthogonal computational bases of the readout qubit. In this case the quantum model is able to map the encoded data from different classes to different orthogonal  bases where the Hilbert space distance is maximal.

An alternative training approach is to use a loss function in terms of a $l_2$ distance metric, or equivalently the HS distance, $\mathcal{D}_{hs}$, as 

\begin{equation}
    \mathcal{D}_{hs}(\rho,\sigma) = \text{Tr}\left[ (\rho-\sigma)^2 \right]\,,
    \label{eq:8}
\end{equation}
where $\rho$ and $\sigma$  are the mixed density matrices sampled from training data of each class, $S$ and $B$, as 

\begin{equation}
    \rho = \frac{1}{M_s}\sum_{s\in S}|s\rangle\langle s| \hspace{6mm} \text{and} \hspace{6mm} \sigma = \frac{1}{M_b}\sum_{b\in B}|b\rangle\langle b|\,,
\end{equation}
where $s$ and $b$ are data points sampled from the training set of each class. The loss function based on the  HS distance can be defined as 
\begin{eqnarray}
    \mathcal{L}(\rho,\sigma)  &=& 1- \frac{1}{2}\mathcal{D}_{hs}(\rho,\sigma) \label{loss:HS} \\
    &=& 1- \frac{1}{2}[\text{Tr}(\rho^2) + \text{Tr}(\sigma^2)] + \text{Tr}(\rho\sigma)\,. \nonumber
\end{eqnarray}
The HS  loss has the advantage that it maximizes the purity of the measured density matrix from each class individually and in the same time it minimizes the overlap between them. Moreover, the HS loss can be minimized in an unsupervised way (without labels). 

Although, the HS loss is an unsupervised loss but it is more efficient than $\mathcal{L}(\vec{\theta},\vec{x})$. The reason is that, $\mathcal{L}(\vec{\theta},\vec{x})$ tries to map the data from each class as close as possible to the corresponding state, i.e. $|0\rangle$ or $|1\rangle$; while $\mathcal{L}(\rho,\sigma)$  maximizes the purity of the measured state from each class by mapping them in different orthogonal bases and in the same time it minimizes the overlap between them  by minimizing the term $\text{Tr}(\rho\sigma)$. 

 In order to evaluate the model classification performance we use a fidelity classifier \cite{Wang:2021shr}. A fidelity classifier is defined as the difference  inner product squared between the embedded test sample $|x\rangle$ and the respective  class, $S$ or $B$, encoded by the training samples, $|s\rangle$ and $|b\rangle$, as \cite{buhrman2001quantum,lloyd2020quantum}
\begin{equation}
    \mathcal{F}(x) = \frac{1}{M_s}\sum_{s\in S} \left|\langle x| s\rangle \right|^2 - \frac{1}{M_b}\sum_{b\in B} \left|\langle x| b\rangle \right|^2\,.
    \label{eq:fid}
\end{equation}
For binary classification, the fidelity classifier assign a binary predicted labels to the input data according to 
\begin{equation}
\hat{Y}  = 
\begin{cases}
            -1 \text{\  if \ } \mathcal{F}(x) < 0 \\
              +1 \text{\  if \ } \mathcal{F}(x) \ge 0
    \end{cases} .
\end{equation}

\section{\bf {Physics analysis}}  
We utilize the quantum metric learning to probe the signature of the CP-odd scalar $(A)$ from the Two Higgs Doublet Model with type II Yukawa structure at the LHC.  We first carry out the analysis using the quantum learning methods with HS loss function  using a single qubit and four qubits circuit as schematically shown in figure \ref{fig:QML}. Also, we examine the expressive power of the VQC with four entangled qubits. Finally, we compare the result with the classical contrastive learning in appendix.\,\ref{sec:contra} when  using JetCLR \cite{Dillon:2021gag}. We consider the process of $pp\to A\to Zh$ which later decay to leptons and b-jet pairs, $Z\to \ell \bar \ell$, and $h\to b\bar b$. For backgrounds, we take $pp\to Zb\bar b$, $pp\to t\bar t$ with a dilepton channel, and $pp\to ZZ(h)$ in the Standard Model process. 
 
We use MadGraph \cite{Alwall:2014hca,Stelzer:1994ta} for Monte Carlo events generation. For the simulation of extra hard radiated jets we use MLM matching \cite{Mangano:2006rw}. Pythia \cite{Sjostrand:2019zhc,Sjostrand:2014zea} is used for parton showering, hadronization, heavy flavor decays and the simulation of  soft underlying events. FastJet \cite{Cacciari:2011ma} is used for jet clustering. For fast detector simulation Delphes package \cite{Ovyn:2009tx} is utilized.  Finally, for b-jet tagging we use a  flat b-tagging efficiency of $70\%$, while for the miss-tagging rate of gluon and light quark jets  we adopt a flat rate of $10^{-3}$.
For event selection cut we consider, (1)  At least two b tagged jets $j_b$ with $R=0.4$ with $P_T(j_{b}) > 45\,\rm{GeV}$ , (2) Two or more isolated leptons, $\ell=\mu, e$, with $P_T(\ell)> 25\,\GeV$, (3) Mass window for the $125\,\rm{GeV}$ reconstructed Higgs as $110 < M_{j_{b}j_{b}} < 140\,\GeV$, (4) Mass window for the reconstructed $Z$ boson as $70 < M_{\ell \bar \ell } < 110\,\rm{GeV}$, (5) Transverse momentum of the b-jet pair has to be $P_T(j_{b}j_{b})> 250\,\GeV$. 
Cut efficiency for signal is about $15\%$ and for the Standard Model backgrounds are $\mathcal{O}(1)\%$.
Considering the selection cuts to maximize the signal to background yield, the dominant background contribution comes from $pp\to Zb\bar b$ with $Z$ boson decays to di-lepton and the leptonic decays for the $tt$ process. Other processes as di-gauge boson production and  $pp\to Zh$  can be easily removed with the basic initial cuts\footnote{ Although the di-gauge boson process has an approximate cut efficiency as the other background processes but it has a smaller cross section of $\sim 0.2$ pb.}. 
After applying the set of cuts we prepare the background set by stacking the events of  $pp\to Z b\bar b$  and $pp\to t\bar t$ according to the cut efficiency times total cross section at $14\,\rm{TeV}$, which is $68\,\rm{pb}$ and $23\,\rm{pb}$ for the two channels respectively.

\subsection{\bf {Data preprocessing}} 
In order to explore the global color structure of the signal and background events, we can use the energy deposit in the calorimeter to depict the information of the final state hadrons into images \cite{Komiske:2016rsd,Fraser:2018ieu,Cogan:2014oua,deOliveira:2015xxd,Esmail:2023axd,Kim:2019wns,Huang:2022rne}. The pixels intensity of the image can be weighted according to the total transverse momentum of the final state hadrons, while the image dimensions will be the pseudorapidity and the azimuth angle of the calorimeter.
For each event we construct an image as a square array in the ($\eta-\phi$) plane with each pixel given by the total hadrons $p_T$ deposited in the associated region in the calorimeter. The rectangular region between $-2.5 \le \eta \le 2.5$ and $-\pi \le \phi \le\pi $  is discretized into $50 \times 50$ pixels grid. This ``jet image" is pre-processed with following steps:
\begin{itemize}
\item \textbf{Image cleaning:}  Images are constructed only from hadrons which have track information. In the meantime we  remove leptons and photons.
\item \textbf{Centering:} Center all particles in the image by shifting  $\left( \frac{\eta_{\bar{b}}+\eta_{b}}{2},\frac{\phi_{\bar{b}}+\phi_{b}}{2} \right)$ to $(0,0)$.  
\item \textbf{Normalization:}  Normalize the pixel intensity by dividing each pixel in the image by the maximum pixel intensity value.
\item \textbf{Pixelization:}  The region in the $(\eta, \phi)$ plane is discretized  into a $50\times 50$ grid with each pixel weighted by the sum of the transverse momentum in it.
\end{itemize}

\begin{figure}[!h]
    \centering
    \includegraphics[width=0.5\textwidth]{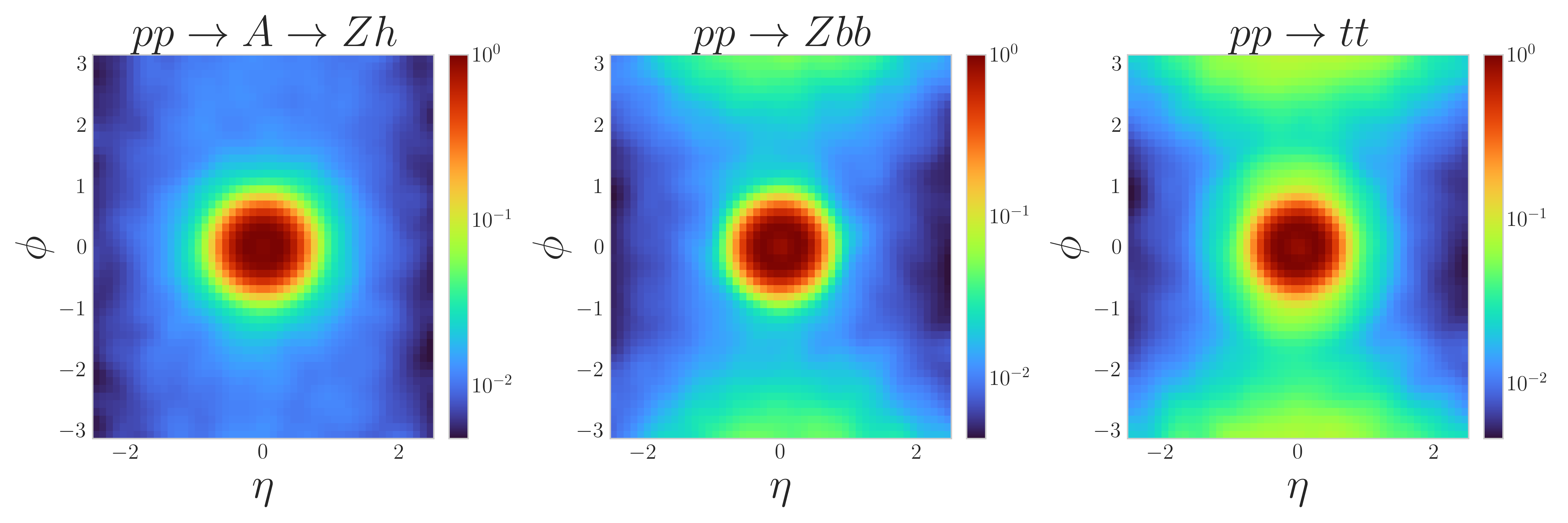}
    \caption{Normalized  $P_T$ distribution of accumulated 50,000 events for the signal and backgrounds.}
    \label{fig:jet_images}
\end{figure}

After pre-processing steps jet images have the dimension of $(50\times 50\times 1)$. In figure 
 \ref{fig:jet_images} we show  a normalized distribution of accumulated $50\,000$ images for the considered signal benchmark and the relevant backgrounds.

As  the input  for JetCLR has a different  structure from jet images, specifically a data set of particle clouds, we preprocess the input data to JetCLR with all the mentioned preprocessing steps except the pixelization step. Moreover, we constrain  the number of charged hadrons in each event to 100, ordered by their transverse momentum,  while events with lower number of constituents are padded by zeros.  After preparing the input data to the different ML models we split the data into $70:30$ for training and testing using equal size data sets of the signal and the backgrounds.  
\begin{figure*}[ht!]
    \includegraphics[width=0.85\textwidth]{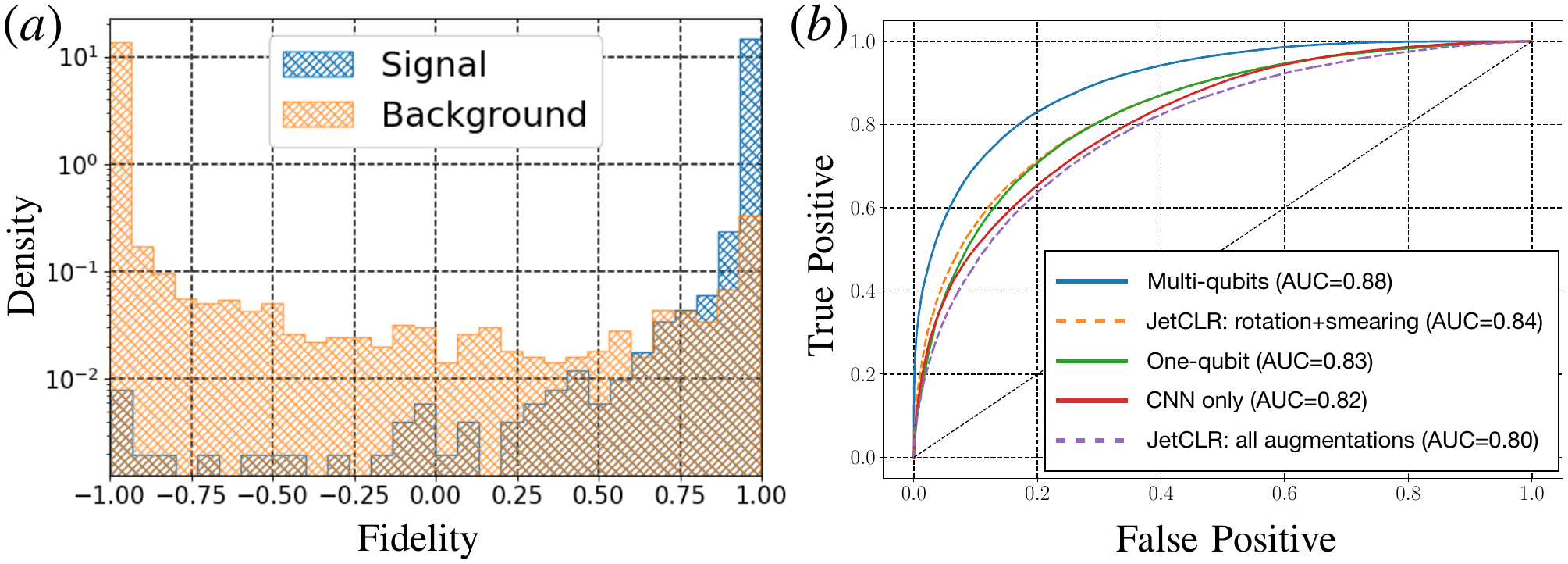}
    \caption{(a) Fidelity  distribution, as defined in eq.\,(\ref{eq:fid}), for the multi-qubits VQC. (b) ROC curves for the classical and quantum models. }
    \label{fig:results}
\end{figure*}
\subsection{\bf {Networks structure}}  
Starting with the  structure of the VQC model, and in order to analyze the jet images we need a VQC with a number of registered qubits equals to the number of pixels in the jet images. As the current and near-term quantum devices suffer from, short  life time of the qubit and small number of entangled qubits  limits our ability to use VQC ``alone" to analyze the jet images. On the other hand, we can adopt a hybrid classical-quantum model which is first introduced in \cite{lloyd2020quantum}. In this case we use the classical CNN model to reduce the dimensions of the jet images before we feed them into the VQC. The classical CNN part consists of an input layer and three couples of convolution layers with sizes of $64,32$ and $16$ and a common kernel of dimensions $3\times 3$. Each convolution pair is  followed by a max-pooling layer with kernel of size $2$ and dropout layer with dropout rate of $20\%$.  The output from the last convolution layer is flattened and followed by a $128$ fully connected layer and one output layer with $3$ neurons, which is the input to the VQC. As the quantum gates used in the VQC consist of asymptotic functions, we normalize the weights of the final linear layer to the unit vector. In this analysis we adjusted two VQCs with the same structure  but different number of registered qubits.  
The structure of the single qubit VQC consists of a data embedding gate and one generic unitary rotation gate which both are repeated $10$ times. The generic unitary gates consist of three parameters that are optimized during the training while the embedding gates are fixed with no trainable parameters. The measurement from the VQC  is the overlap between the measured states with the pure computational states of the qubit. To exploit the advantage of the qubits interference, we increase the depth of the VQC with $4$ entangled qubits  via the two qubits controlled-$Z$ rotation gate. In this case one can think of each individual qubit as a neural network layer while the number of the entangled qubits represents the depth of the network. Both data encoding gates and the unitary rotation gates are repeated for $6$ times. Although we can measure the overlap on all the four qubits,  we opted to follow the methods in \cite{perez2020data} with only one readout qubit as shown in figure \ref{fig:QML}. This setup enables us to use the same training procedures as in the single qubit classifier. A schematic architecture of the model is shown in figure \ref{fig:QML}. Input data  is prepared as pairs of images with size $10$ K  pair for training and  $15\, 000$  pair for test. We train the model for $50$ epochs with a batch of size $10$ and Adam optimizer is used to update the weights of the classical and VQC together. For the quantum  simulation we use PennyLane Frame work \cite{bergholm2018pennylane} together with  TensorFlow \cite{tensorflow2015-whitepaper}.

For JetCLR setup we use an input data of dimensions (Events, Number of constitutes, features) with features considered as $(P_T,\eta,\phi)$ and maximum number of constitutes of $100$ while events with lower number are padded with zeros such that the maximum length is retained. It is worth mentioning that we order the constituents of each event in a $P_T$ descending order to avoid missing any physics for events with larger event constitutes. We use the publicly available  package in \href{https://github.com/bmdillon/JetCLR} {JetCLR}, with four transformer layers each consists of four self attention heads and  a projection head of dimension $1000$.  For the first training stage we minimize a contrastive loss function with regularization temperature parameter of $0.1$ and training for $100$ epochs using Adam optimizer. The second training step is done with a linear fully connected layer of dimension $1000$ and trained for $750$ epochs using Adam optimizer with $10^{-3}$ learning rate. The model is trained on $100,000$ events for the signal and background events. It is important to mention that JetCLR is a semi-supervised training  model which has no labels in the first training stage  but with some pseudo labels in the second one.

\begin{figure*}[t!]
    \centering
    \includegraphics[width=0.85\textwidth]{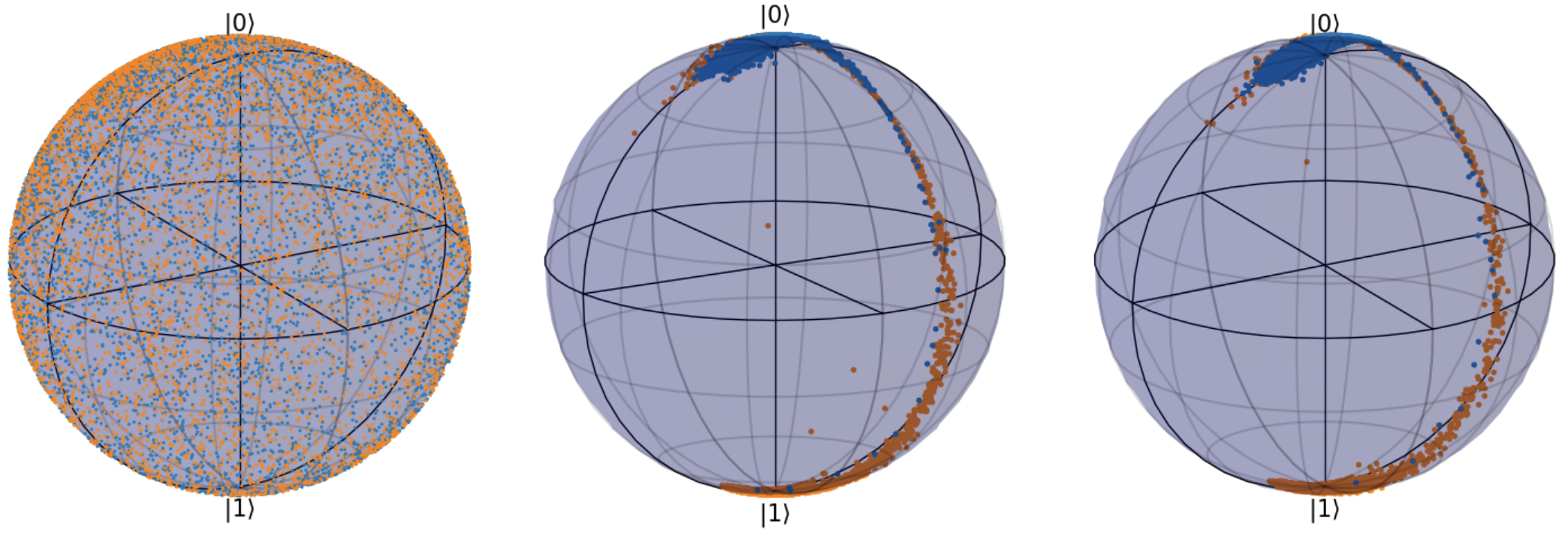}
    \caption{Bloch sphere representation of the readout qubit with signal (blue points) and background events (orange points). Left: Bloch sphere representation of the  training data at the first iteration; Middle: Representation of the training data after 50 epochs; Right: Representation of the test data. Please note that the arc like structure of the data as we normalize the VQC input to a unit vector.}
    \label{fig:qubits}
\end{figure*}
\subsection{\bf {Result}} 

After applying  different classical and quantum models to prob the color structure of the process $pp\to A\to hZ \to \bar{l}l\bar{b}b$ from the relevant background, we quantify the classification performance of all used models by reporting the ROC curve. 
The output of the quantum model is shown in figure \ref{fig:results}(a) in which the  events with positive fidelity are classified as most likely signal events while events with negative fidelity are most likely background events. To quantify the classification performance 
the Area Under the Curve (AUC)  measures the ability of a binary classifier to distinguish between classes and is used as a summary of the ROC curve. In figure \ref{fig:results}(b) we show the ROC curves for all used self-supervised learning models. The  VQC with classical encoder model shows the best performance among all other models with AUC = $0.88$. 

JetCLR results with all augmentations, i.e. Rotation, translation, momentum smearing and collinear splittings, shows lower performance with AUC = $0.80$, as the translation and collinear augmentations wash out the characteristic features between the signal and background events which in turn hinders the classification performance. On the other hand, JetCLR with only rotation and  momentum smearing augmentations exhibits a larger classification performance with AUC= $0.84$. As mentioned previously,  one of the major drawbacks of classical contrastive learning is that applying data augmentations with equal magnitude may hinder the classification performance.  

VQC with a single qubit has AUC = $0.83$. We stress here that hyper-parameters of all the used models have been optimised to its best classification performance individually. 
To assess the result of the quantum circuit, figure \ref{fig:qubits} shows the Bloch sphere representation of the readout qubit  with signal images as blue points and background images as red points. At the first iteration we randomly initialize the weights of the classical CNN and VQC which exhibit a random distribution of the data on the readout qubit. After training the model for $50$ epochs, the middle qubit shows a well distributed points  where the signal images clustered onto the $|0\rangle$ state and background images cluster around $|1\rangle$ state. To test the generalization of the quantum model, the right qubit shows the distribution of the test images. We stress here that the  arc shape of the distributed points  is due to the normalization of the input weights to unity.

\section{\bf {Conclusion}}
In this paper we check a possibility to utilize ``Quantum advantage" in analysing collider data. For a concrete example, we consider a CP-odd scalar signature at the LHC against irreducible QCD backgrounds. Thus we need to utilize the peculiar QCD color structure of the signal with respect to the background as a discriminating variable to distinguish the signal from the background events. Since the energy deposits from QCD activities are soft and locate sparsely across various sub-detectors, this will bring a similar problem like the curse of dimensionality, i.\,e.\, difficulty in finding correlations among data points. To resolve this difficulty, we adopt different semi-supervised classical and quantum machine learning models based on metric learning.  We first adjusted the JetCLR model which is a semi-supervised twin encoders model that learns the similarity and dissimilarity between the input data by mapping them into different regions in the latent space. To do so, the model relies on data augmentations to generalize the latent space boundaries for different classes. We found that applying different data augmentations equally, Rotation, translation, momentum smearing and collinear splittings, hinders the classification performance leading to AUC = $80\%$ while including rotation and momentum smearing only, enhances the classification performance with AUC=$84\%$. In fact, classical contrastive learning models can efficiently learn a representative structure of the input data with higher efficiency  as far as the data augmentations impact the classification boundaries in the latent space \cite{Dillon:2021gag}.  
In the considered analysis we found that data augmentations do not affect the classification accuracy and thus we adopted an alternative hybrid classical-quantum approach based on quantum metric learning. The hybrid model can classify the input data, with out labels, by mapping the data from different classes onto different orthogonal bases of the readout qubit. The hybrid model with four registered qubits shows an improvement over the classical contrastive learning model. 

Finally, it is crucial to emphasize that the results obtained for the quantum models stem from a classical simulation, wherein we disregard the decoherence effect. When tested on an actual quantum computer, we anticipate a slight reduction in classification performance; however, we remain optimistic that the impact of quantum noise will be significantly diminished in the near future, allowing us to recover and validate our results more accurately.

\section*{Acknowledgments}
MP is supported by National Research Foundation of Korea, NRF-2021R1A2C4002551.
AH is funded by the grant number 22H05113, ``Foundation of Machine Learning Physics", Grant in Aid for Transformative Research Areas and 22K03626, Grant-in-Aid for Scientific Research (C). KK is supported  by the US Department of Energy under Grant no.\,DE-SC0024673.

\appendix
\section{\bf {Classical metric (contrastive) learning}}{\label{sec:contra}} 
Classical Contrastive learning has been proved to be able to learn meaningful visual representations without labels.  Originally, contrastive method  aims to classify data by creating two views of the data from the same samples, by augmentation such as rotation, crop, translation, etc, as positive pair while the views from different samples as negative pair. To do so, contrastive learning has a two steps of training;  In the first  step the model minimizes the distance between  positive pairs and maximizes the distance between the negative pairs in the latent space of the model using a contrastive loss defined as \cite{khosla2020supervised}    
\begin{equation}
    \mathcal{L}_i = - \log  \frac{e^{s(z_i,z_{i}')/\tau} }{\sum_{j \neq i \in \text{batch}}^{} \left[ e^{s(z_i,z_{j}')/\tau} \right] } \,,
    \label{eq:contrastive}
\end{equation}
where $s(z_i,z_j)$ is the cosin similarity between  pairs from the samples $i$ and $j$ respectively; primed $z_i$ is the augmented version of the sample. Once the distance in the latent space between the data from different classes is maximal, the second training step assumes pseudo labels and adds one fully connected layer to classify the mapped data. It is seen that  the important  process in contrastive learning is the data augmentation which is essential to minimize the contrastive loss. In fact, the
main principle of contrastive learning is to learn representational invariance by making the network  invariant under a set of data augmentations \cite{tian2020makes}. Thus if the augmented data do not exhibit a large difference between samples from different classes the classification boundaries on the latent space will not be identified efficiently which in turn hinders the classification performance.    

\bibliographystyle{unsrt}
\bibliography{main}
\end{document}